\documentstyle[12pt]{article}
\begin{document}
\tolerance=5000
\def\be{\begin{equation}}
\def\ee{\end{equation}}
\def\bea{\begin{eqnarray}}
\def\eea{\end{eqnarray}}
\def\nn{\nonumber \\}
\def\cF{{\cal F}}
\def\det{{\rm det\,}}
\def\Tr{{\rm Tr\,}}
\def\e{{\rm e}}
\def\etal{{\it et al.}}
\def\erp2{{\rm e}^{2\rho}}
\def\erm2{{\rm e}^{-2\rho}}
\def\er4{{\rm e}^{4\rho}}
\def\etal{{\it et al.}}

\  \hfill 
\begin{minipage}{3.5cm}
OCHA-PP-147 \\
NDA-FP-69 \\
\end{minipage}

\vfill

\begin{center}
{\large\bf Conformal Anomaly from d5 Gauged Supergravity and 
c-function Away from Conformity}

\vfill

{\sc Shin'ichi NOJIRI}\footnote{nojiri@cc.nda.ac.jp}, 
{\sc Sergei D. ODINTSOV}$^{\spadesuit}$\footnote{
odintsov@mail.tomsknet.ru}, \\
{\sc Sachiko OGUSHI}$^{\heartsuit}$\footnote{
JSPS Research Fellow, 
g9970503@edu.cc.ocha.ac.jp}
\\

\vfill

{\sl Department of Mathematics and Physics \\
National Defence Academy, 
Hashirimizu Yokosuka 239, JAPAN}

\vfill

{\sl $\spadesuit$ 
Tomsk Pedagogical University, 634041 Tomsk, RUSSIA}

\vfill

{\sl $\heartsuit$ Department of Physics, 
Ochanomizu University \\
Otsuka, Bunkyou-ku Tokyo 112, JAPAN}

\vfill

{\bf ABSTRACT}

\end{center}
Using AdS/CFT correspondence we found the conformal anomaly from
d3 and d5 gauged supergravity with single scalar (dilaton) and the arbitrary 
scalar potential on AdS-like scalar-gravitational background. Such
dilatonic gravity action describes the special RG flows 
in extended gauged SG when scalars lie in one-dimensional submanifold of
complete scalars space. This dilaton-dependent conformal anomaly 
corresponds to dual non-conformal (gauge) QFT (which is classically
conformally invariant) with account of radiative
corrections. Equations of motion in d5 gauged supergravity put some
restrictions to the dilatonic potential on the conformal boundary.
Using these restrictions we propose the candidate c-functions away 
from exact conformity. These c-functions are positively defined and
monotonic, expressed in terms of dilatonic potential and have the fixed 
points in asymptotically AdS region.

\newpage

Five-dimensional gauged supergravity (SG) plays an important role in
AdS/CFT correspondence\cite{AdS}. It is known that different versions 
of d5 gauged SG (for example, ${\cal N}=8$ d5 gauged SG\cite{GRW} 
with fourty-two scalars and non-trivial scalar potential) may appear 
as a result of truncation of d10 IIB SG. In particular, 
${\rm AdS}_5\times{\rm S}_5$ deformed 
truncation of IIB SG (with non-trivial scalars) corresponds to some 
specific solution of d5 gauged SG. Hence, it is often enough to study 
5d gauged SG classical solutions in AdS/CFT set-up instead of the
investigation of 
non-linear IIB SG solutions. Such (deformed) solutions describe 
RG flows in dual boundary field theory (for a very recent discussion 
of such flows, see \cite{GPPZ,CM,DF} and refs. therein). 
It is very interesting that even 4d curvature or non-zero 
temperature effects may be taken into 
account in bulk description of such RG flows \cite{NOR}. 
In consideration of extended d5 gauged SG solutions there are often 
more symmetric (special) 
RG flows where scalars lie in one-dimensional submanifold of complete 
scalars space. (Then such theory corresponds to d5 dilatonic 
gravity with non-trivial dilaton potential). 
Such flows may also correspond to certain (D3)-brane
distributions \cite{FGPW}.
However, note that it is extremely difficult to make the 
explicit identification of deformed gauged SG solution with
the corresponding non-conformal dual gauge theory\footnote{
By this we mean that in flat space only ${\cal N}=4$ SYM theory is 
considered to be exactly conformally invariant one among interacting 
theories. As a result, its exact conformal anomaly which is not
renormalized is known. Of course, all dual boundary theories 
we consider, like QCD, etc. are classically conformally invariant 
in curved spacetime, 
while we call them here non-conformal to mention the difference 
with ${\cal N}=4$ SYM. Hence, all these theories have well-defined 
conformal anomalies with account of radiative
corrections. However, these conformal anomalies are 
explicitly unknown. Their calculation has been done in 
few simple theories (QED, gauge theory
without fermions) but only up to two- or three-loops.
It was shown \cite{BCH} that such radiatively corrected CA 
contains not only $F$, $G$ terms but also $R^2$. 
It is a challenge to find exact conformal anomaly for 
gauge theories. Probably, only SG-description may help 
in resolution of this problem.}.

The important characteristic of boundary (gauge) theory in AdS/CFT 
correspondence is the conformal anomaly which may be found from 
the bulk side (see paper by Witten in ref.\cite{AdS}). 
The calculation of conformal anomaly 
in d5 gauged SG with single scalar and constant scalar potential 
(dilatonic gravity) on dilaton-gravitational background via AdS/CFT 
correspondence has been initiated in ref.\cite{NOano}. It was 
shown that ${\cal N}=4$ super YM 
theory covariantly coupled with ${\cal N}=4$ conformal 
SG  \cite{peter} is actual dual 
of d5 dilatonic gravity (see also derivation of anomaly in 
gravity-complex scalar background in refs.\cite{LT,NOOSY}). 
 From holographic RG description (see refs.\cite{BK,VV} for 
introduction) 
it is known that dilaton (or in more general case, scalars) describe 
the couplings of dual (gauge) theory, say, masses, 
scalars or coupling constants.
Hence, it is extremely interesting to get the conformal anomaly 
from gauged SG with non-trivial scalar potential. 
This may give much better understanding 
of RG flows in dual (non-conformal exactly) boundary theory and also 
the definition of analog of central charge (c-function) away of exact
conformity. 
Even more, considering 
the conformal anomaly (up to some loop) of dual general boundary theory
with radiative 
corrections 
and comparing it with the one from bulk gauged SG may help 
in correct identification of dual boundary theory with 
correspondent bulk identification (which is currently
non-easy task). Note also that conformal anomaly plays an important 
role in
the construction of local surface counterterm for gauged SGs with
non-constant scalar potential \cite{CO}.

In the present letter we find the AdS/CFT conformal anomaly from d3 
and d5 gauged SG with single scalar (dilaton) and arbitrary dilaton 
potential.
This situation corresponds to special RG flow in dual description.
The acceptable candidates for analogs of central charge (or more exactly, 
of c-function) away of exact conformity 
are proposed when equations of motion are partly used.
 The monotonity and positivity of c-functions is shown,
they have standard fixed points in asymptotically AdS region. 

We start with the bulk action of $d+1$-dimensional 
dilatonic gravity with the potential $\Phi $
\be
\label{i}
S={1 \over 16\pi G}\int_{M_{d+1}} d^{d+1}x \sqrt{-\hat G}
\left\{ \hat R + X(\phi)(\hat\nabla\phi)^2 
+ Y(\phi)\hat\Delta\phi
+ \Phi (\phi)+4 \lambda ^2 \right\} \ .
\ee
Here $M_{d+1}$ is  $d+1$ dimensional manifold whose 
boundary is the $d$ dimensional manifold $M_d$ and 
we choose $\Phi(0)=0$. Such action corresponds to
(bosonic sector) of gauged SG with single scalar (special RG flow).
Note also that classical vacuum stability restricts the form of 
dilaton potential \cite{T}.
As well-known, we also need to add the surface terms \cite{3} 
to the bulk action in order that the variational principle 
to be well-defined.
We should only note here that the surface terms become 
irrelevant finally in the calculation for the Weyl anomaly 
given in this work.
The equations of motion given by variation of (\ref{i}) with
respect to $\phi$ and $G^{\mu \nu}$ are  
\bea
\label{ii}
0&=&-\sqrt{-\hat{G}}\Phi'(\phi)-\sqrt{-\hat{G}}V'(\phi)
{\hat G}^{\mu \nu}\partial_{\mu }\phi 
\partial_{\nu}\phi \nn
&& +2 \partial_{\mu }\left(\sqrt{-\hat{G}}
\hat{G}^{\mu \nu}V(\phi)\partial_{\nu} \phi \right) \\
\label{iii}
0 &=& {1 \over d-1}\hat{G}_{\mu\nu}\left(
\Phi(\phi)+{d(d-1) \over l^2}\right)+\hat{R}_{\mu \nu}+V(\phi)
\partial_{\mu }\phi\partial_{\nu}\phi \ .
\eea
Here $V(\phi)\equiv X(\phi) - Y'(\phi)$.

We choose the metric $\hat G_{\mu\nu}$ on $M_{d+1}$ and 
the metric $\hat g_{\mu\nu}$ on $M_d$ in the following form
\be
\label{ib}
ds^2\equiv\hat G_{\mu\nu}dx^\mu dx^\nu 
= {l^2 \over 4}\rho^{-2}d\rho d\rho + \sum_{i=1}^d
\hat g_{ij}dx^i dx^j \ ,\quad 
\hat g_{ij}=\rho^{-1}g_{ij}\ .
\ee
Here $l$ is related with $\lambda^2$ 
by $4\lambda ^2 = d(d-1)/{l^{2}}$.
If $g_{ij}=\eta_{ij}$, the boundary of AdS lies at $\rho=0$. 
Note that we follow the method of calculation 
in refs.\cite{NOano,NOOSY} 
where dilatonic gravity with constant dilaton potential 
has been considered. 

The action (\ref{i}) diverges in general since the action 
contains the infinite volume integration on $M_{d+1}$. 
The action is regularized by introducing the infrared cutoff 
$\epsilon$ and replacing 
\be
\label{vibc}
\int d^{d+1}x\rightarrow \int d^dx\int_\epsilon d\rho \ ,\ \ 
\int_{M_d} d^d x\Bigl(\cdots\Bigr)\rightarrow 
\int d^d x\left.\Bigl(\cdots\Bigr)\right|_{\rho=\epsilon}\ .
\ee
We also expand $g_{ij}$ and $\phi$ with respect to $\rho$: 
\be
\label{viib}
g_{ij}=g_{(0)ij}+\rho g_{(1)ij}+\rho^2 g_{(2)ij}+\cdots \ ,\quad
\phi=\phi_{(0)}+\rho \phi_{(1)}+\rho^2 \phi_{(2)}+\cdots \ .
\ee
Then the action is also expanded as a power series on $\rho$. 
The subtraction of the terms proportional to the inverse power of 
$\epsilon$ does not break the invariance under the scale 
transformation $\delta g_{ \mu\nu}=2\delta\sigma g_{ \mu\nu}$ and 
$\delta\epsilon=2\delta\sigma\epsilon$ . When $d$ is even, however, 
the term proportional to $\ln\epsilon$ appears. This term is not 
invariant under the scale transformation and the subtraction of 
the $\ln\epsilon$ term breaks the invariance. The variation of the 
$\ln\epsilon$ term under the scale transformation 
is finite when $\epsilon\rightarrow 0$ and should be canceled 
by the variation of the finite term (which does not 
depend on $\epsilon$) in the action since the original action 
(\ref{i}) is invariant under the scale transformation. 
Therefore the $\ln\epsilon$ term $S_{\rm ln}$ gives the Weyl 
anomaly $T$ of the action renormalized by the subtraction of 
the terms which diverge when $\epsilon\rightarrow 0$ (d=4) 
\be
\label{vib}
S_{\rm ln}=-{1 \over 2}
\int d^4x \sqrt{-g }T\ .
\ee

First we consider the case of $d=2$, i.e. three-dimensional gauged SG.
The anomaly term $S_{\rm ln}$ proportional 
to ${\rm ln}\epsilon$ in the action is 
\bea
\label{IIii}
S_{\rm ln}=-{1 \over 16\pi G}{l\over 2}\int d^{2}x 
\sqrt{-g_{(0)}}
\left\{ R_{(0)} + X(\phi_{(0)})(\nabla\phi_{(0)})^2 + Y(\phi_{(0)})
\Delta\phi_{(0)} \right.&&\nn
\left. + \phi _{(1)}\Phi '(\phi_{(0)})+{1 \over 2}
g^{ij}_{(0)}g_{(1)ij}\Phi(\phi_{(0)}) \right\} . && 
\eea
The terms propotional to $\rho ^{0}$ with $\mu ,\nu =i,j $  
in (\ref{iii}) lead to $g_{(1)ij}$ in terms of $g_{(0)ij}$ and
$\phi_{(1)}$.
\bea  
\label{vi}
g_{(1)ij}&=&\left[-R_{(0)ij}-V(\phi_{(0)})
\partial_i\phi_{(0)}\partial_j\phi_{(0)}
 -g_{(0)ij}\Phi'(\phi_{(0)})\phi_{(1)}\right.\nn
&& +{g_{(0)ij} \over l^2}\left\{2\Phi'(\phi_{(0)})\phi_{(1)}
+R_{(0)}+V(\phi_{(0)})g_{(0)}^{kl}\partial_k\phi_{(0)}
\partial_l\phi_{(0)}  \right\}\nn
&& \times \left.\left( \Phi(\phi_{(0)})
+{2 \over l^2} \right)^{-1}\right] 
\times \Phi(\phi_{(0)})^{-1} 
\eea
In the equation (\ref{ii}), the terms propotional to $\rho ^{-1}$
lead to $\phi_{(1)}$ as follows: 
\bea
\label{vii}
\phi_{(1)}&=& \left[  V'(\phi_{(0)}) 
g_{(0)}^{ij}\partial_i\phi_{(0)}\partial_j\phi_{(0)} 
+ 2 {V(\phi_{(0)})  \over \sqrt{-g_{(0)}}} \partial_i
\left(\sqrt{-g_{(0)}}
g_{(0)}^{ij}\partial_j\phi_{(0)} \right) \right.\nn 
&& \left. +{1 \over 2}\Phi'(\phi_{(0)})
\left( \Phi(\phi_{(0)})+{2 \over l^2} \right)^{-1}
\{ R_{(0)}+V(\phi_{(0)})
g^{ij}_{(0)}\partial_i\phi_{(0)}\partial_j\phi_{(0)} \} \right] \nn
&& \times \left( \Phi ''(\phi_{(0)})
 -\Phi'(\phi_{(0)})^2  \left( \Phi(\phi_{(0)})
+{2 \over l^2} \right)^{-1} \right)^{-1}
\eea
Then anomaly term takes the following form  using
(\ref{vi}), (\ref{vii})
\bea
\label{ano2}
S_{\rm ln}&=&-{1 \over 16\pi G}{l\over 2}\int d^{2}x 
\sqrt{-g_{(0)}}
\left\{  R_{(0)}+X(\phi_{(0)})(\nabla\phi_{(0)})^2 + Y(\phi_{(0)})
\Delta\phi_{(0)} \right. \nn
&& +{1 \over 2}\left\{ {3 \Phi '(\phi_{(0)}) \over l^2 }
\left(\Phi ''(\phi_{(0)})
\left( \Phi(\phi_{(0)})+{2 \over l^2 }\right)
 -\Phi'(\phi_{(0)})^2\right)^{-1} -\Phi(\phi_{(0)})\right\}  \nn
&& \times \left(R_{(0)} +V(\phi_{(0)})g^{ij}_{(0)}\partial_i\phi_{(0)}
\partial_j\phi_{(0)} \right)\left( \Phi(\phi_{(0)})
+{2 \over l^2 }\right)^{-1} \nn
&& +{2 \Phi '(\phi_{(0)}) \over l^2 }
\left(\Phi ''(\phi_{(0)})
\left( \Phi(\phi_{(0)})+{2 \over l^2 }\right)
 -\Phi'(\phi_{(0)})^2\right)^{-1} \nn
&& \left. \times \left(V'(\phi_{(0)}) 
g_{(0)}^{ij}\partial_i\phi_{(0)}\partial_j\phi_{(0)} 
+ 2 {V(\phi_{(0)})  \over \sqrt{-g_{(0)}}} 
\partial_i\left(\sqrt{-g_{(0)}}
g_{(0)}^{ij}\partial_j\phi_{(0)} \right) \right) \right\}\ .
\eea
For $\Phi(\phi)=0$ case, the central charge of the conformal 
field theory is given by the coefficient of $R$. Then it 
might be natural to introduce the analog of central charge $c$, i.e.
c-function for the case 
when the 
conformal symmetry is broken by the deformation as follows :
\bea
\label{d2c}
c&=&{1 \over 2G}\left[ l +{l \over 2}\left\{ {2 \Phi '(\phi_{(0)}) 
\over l^2 } \left(\Phi ''(\phi_{(0)})
\left( \Phi(\phi_{(0)}) \right.\right.\right.\right. \nn
&& \left.\left.\left.\left. +{2 \over l^2 }\right)
 -\Phi'(\phi_{(0)})^2\right)^{-1} -\Phi(\phi_{(0)})\right\} 
\times \left( \Phi(\phi_{(0)})
+{2 \over l^2 }\right)^{-1}  \right]\ .
\eea
Comparing this with radiatively-corrected c-function of boundary 
QFT may help in correct bulk description of such theory.
If candidate c-function is getting non-monotonic in some region that
indicates to breaking of SG description there.

We now consider the case of $d=4$. 
The anomaly terms which proportional to ${\rm ln }\epsilon$
are
\bea
\label{ano}
S_{\rm ln}&=&{1 \over 16\pi G}\int d^4x \sqrt{-g_{(0)}}\left[ 
{-1 \over 2l}g_{(0)}^{ij}g_{(0)}^{kl}\left(g_{(1)ij}g_{(1)kl}
 -g_{(1)ik}g_{(1)jl}\right) \right. \nn
&& +{l \over 2}\left(R_{(0)}^{ij}-{1 \over
2}g_{(0)}^{ij}R_{(0)}\right)g_{(1)ij} \nn
&& -{2 \over l}V(\phi_{(0)})\phi_{(1)}^2
+{l \over 2}V'(\phi_{(0)})\phi_{(1)}
g_{(0)}^{ij}\partial_i\phi_{(0)}\partial_j\phi_{(0)} \nn
&& +l V(\phi_{(0)})\phi_{(1)}
{1 \over \sqrt{-g_{(0)}}}
\partial_i\left(\sqrt{-g_{(0)}}g_{(0)}^{ij}
\partial_j\phi_{(0)} \right) \nn
&&  +{l \over 2}V(\phi_{(0)})\left( g_{(0)}^{ik}g_{(0)}^{jl}
g_{(1)kl}-{1 \over 2}g_{(0)}^{kl}
g_{(1)kl}g_{(0)}^{ij}\right)  \partial_i\phi_{(0)} 
\partial_j\phi_{(0)}  \\
&& - {l \over 2}\left({1 \over 2}g_{(0)}^{ij}g_{(2)ij}
 -{1 \over 4}g_{(0)}^{ij}g_{(0)}^{kl}g_{(1)ik}g_{(1)jl}
+{1 \over 8}(g_{(0)}^{ij}g_{(1)ij})^2 \right)\Phi(\phi_{(0)})\nn
&& \left. -{l \over 2}\left( \Phi'(\phi_{(0)})\phi_{(2)}+
{1 \over 2}\Phi''(\phi_{(0)}) \phi_{(1)}^2 +
{1 \over 2}g_{(0)}^{kl}g_{(1)kl}\Phi'(\phi_{(0)}) \phi_{(1)} \right)
\right]
\ .\nonumber
\eea
The terms proportional to $\rho ^{0}$ with $\mu ,\nu =i,j $  
in the equation of the motion (\ref{iii}) lead to $g_{(1)ij}$ 
in terms of $g_{(0)ij}$ and $\phi_{(1)}$.
\bea  
\label{vibb}
g_{(1)ij}&=&\left[-R_{(0)ij}-V(\phi_{(0)})
\partial_i\phi_{(0)}\partial_j\phi_{(0)}
 -{1 \over 3}g_{(0)ij}\Phi'(\phi_{(0)})\phi_{(1)}\right.\nn
&& +{g_{(0)ij} \over l^2}\left\{{4 \over 3}\Phi'(\phi_{(0)})\phi_{(1)}
+R_{(0)}+V(\phi_{(0)})g_{(0)}^{kl}\partial_k\phi_{(0)}
\partial_l\phi_{(0)}  \right\}\nn
&& \times \left.\left( {1 \over 3}\Phi(\phi_{(0)})
+{6 \over l^2} \right)^{-1}\right] 
\times \left( {1 \over 3}\Phi(\phi_{(0)})
+{2 \over l^2} \right)^{-1} \ .
\eea
In the equation (\ref{ii}), the terms proportional to $\rho^{-2}$
lead to $\phi_{(1)}$ as follows:
\bea
\label{vii4d}
\phi_{(1)}&=& \left[  V'(\phi_{(0)}) 
g_{(0)}^{ij}\partial_i\phi_{(0)}\partial_j\phi_{(0)} 
+ 2 {V(\phi_{(0)})  \over \sqrt{-g_{(0)}}} \partial_i\left(\sqrt{-g_{(0)}}
g_{(0)}^{ij}\partial_j\phi_{(0)} \right) \right.\nn 
&& \left. +{1 \over 2}\Phi'(\phi_{(0)})
\left( {1 \over 3}\Phi(\phi_{(0)})+{6 \over l^2} \right)^{-1}
\{ R_{(0)}+V(\phi_{(0)})
g^{ij}_{(0)}\partial_i\phi_{(0)}\partial_j\phi_{(0)} \} \right] \nn
&& \times \left( {8 V(\phi_{(0)}) \over l^2 } +\Phi ''(\phi_{(0)})
 -{2 \over 3}\Phi'(\phi_{(0)})^2  \left( {1 \over 3}\Phi(\phi_{(0)})
+{6 \over l^2} \right)^{-1} \right)^{-1}\ .
\eea
In the equation (\ref{iii}), the terms proportional to
$\rho^1$ with $\mu ,\nu =i,j$ lead to $g_{(2)ij}$.
\bea
\label{viii}
g_{(2)ij}&=& \left[ -{1 \over 3}\left\{ g_{(1)ij}\Phi'(\phi_{(0)})\phi_{(1)}
+g_{(0)ij}(\Phi'(\phi_{(0)})\phi_{(2)}+{1 \over 2}
\Phi''(\phi_{(0)})\phi_{(1)}^{2} ) \right\}\right.\nn
&& -{2 \over l^2}g^{kl}_{(0)}g_{(1)ki}g_{(1)lj}
+{1\over l^2}g^{km}_{(0)}g^{nl}_{(0)}g_{(1)mn}g_{(1)kl}g_{(0)ij}\nn
&& -{2 \over l^2}g_{(0)ij}\left( {1 \over 3}\Phi(\phi_{(0)})
+{8 \over l^2} \right)^{-1}\times \left\{ {2 \over l^2}
g^{mn}_{(0)}g^{kl}_{(0)}g_{(1)km}g_{(1)ln} \right.\nn
&&-{4 \over 3}\left(\Phi'(\phi_{(0)})\phi_{(2)}+{1 \over 2}
\Phi''(\phi_{(0)})\phi_{(1)}^2 \right) 
 -{1 \over 3}g^{ij}_{(0)}g_{(1)ij}\Phi'(\phi_{(0)})\phi_{(1)}\nn
&& \left.+V'(\phi_{(0)})\phi_{(1)}g^{ij}_{(0)}\partial_i\phi_{(0)}
\partial_j\phi_{(0)}+{2 V(\phi_{(0)})\phi_{(1)} \over \sqrt{-g_{(0)} } }
\partial_i \left( \sqrt{-g_{(0)}}g^{ij}_{(0)}
\partial_j\phi_{(0)}\right) \right\} \nn
&& \left.+V'(\phi_{(0)})\phi_{(1)}\partial_i\phi_{(0)} 
\partial_j\phi_{(0)}+2V(\phi_{(0)})\phi_{(1)}\partial_i
\partial_j\phi_{(0)} \right] \nn
&& \times \left( {1 \over 3}\Phi(\phi_{(0)}) \right)^{-1}\ .
\eea
And the terms proportional to $\rho ^{-1}$ in the equation 
(\ref{ii}), lead to $\phi_{(2)}$ as follows:
\bea
\label{phi2}
\phi_{(2)}&=&\left[ V''(\phi_{(0)})\phi_{(1)}g^{ij}_{(0)}
\partial_i\phi_{(0)}\partial_j\phi_{(0)} \right.\nn
&&+V'(\phi_{(0)})\left( g^{ik}_{(0)}g^{jl}_{(0)}-{1 \over 2}
g^{ij}_{(0)}g^{kl}_{(0)}\right)g_{(1)kl}\partial_i\phi_{(0)}
\partial_j\phi_{(0)} \nn
&&+{2 V'(\phi_{(0)})\phi_{(1)} \over \sqrt{-g_{(0)} } }
\partial_i \left( \sqrt{-g_{(0)}}g^{ij}_{(0)}
\partial_j\phi_{(0)} \right) \nn
&& -{4 \over l^2}V'_{(0)}\phi_{(1)}^2-{1 \over 2}\Phi'''(\phi_{(0)})
\phi_{(1)}^2-{1 \over 2}g^{kl}_{(0)}g_{(1)kl}\Phi''(\phi_{(0)})
\phi_{(1)} \nn
&& -\left({-1 \over 4}g^{ij}_{(0)}g^{kl}_{(0)}g_{(1)ik}g_{(1)jl} 
+{1 \over 8}(g^{ij}_{(0)}g_{(1)ij})^2 \right)\Phi'(\phi_{(0)}) \nn
&& -{1 \over 2}\Phi'(\phi_{(0)})\left( {1 \over 3}\Phi(\phi_{(0)})
+{8 \over l^2} \right)^{-1}\times \left\{ {2 \over l^2}
g^{mn}_{(0)}g^{kl}_{(0)}g_{(1)km}g_{(1)ln} \right.\nn
&& -{2 \over 3}\Phi''(\phi_{(0)})\phi_{(1)}^2 
 -{1 \over 3}g^{ij}_{(0)}g_{(1)ij}\Phi'(\phi_{(0)})\phi_{(1)}\nn
&& \left.\left.+V'(\phi_{(0)})\phi_{(1)}g^{ij}_{(0)}\partial_i\phi_{(0)}
\partial_j\phi_{(0)}
+{2 V(\phi_{(0)})\phi_{(1)} \over \sqrt{-g_{(0)} } }
\partial_i \left( \sqrt{-g_{(0)}}g^{ij}_{(0)}
\partial_j\phi_{(0)}\right)  \right\} \right] \nn
&& \times \left( \Phi''(\phi_{(0)}) -{2 \over 3}\Phi'(\phi_{(0)})^2
\left( {1 \over 3}\Phi(\phi_{(0)})
+{8 \over l^2} \right)^{-1} \right)^{-1}
\eea
Then we can get the anomaly terms (\ref{ano}) in terms of
$g_{(0)ij}$ and $\phi_{(0)}$, which are boundary values of 
metric and dilaton respectively by using (\ref{vibb}), (\ref{vii4d}),
(\ref{viii}), (\ref{phi2}). 
As we are only interested in c-function  
 away from conformity, 
we  present the coefficients of $R_{(0)}^2$ 
and $R_{(0)ij}R_{(0)}^{ij}$, which appear in Weyl anomaly 
 (\ref{ano2}). For this reason, we neglect the terms 
containing the derivative with respect to $x_i$. 
 Here we choose $l=1$ and denote 
$\Phi(\phi_{(0)})$ by $\Phi$ and abbreviate the index $(0)$ 
for the simplicity.

Then substituting (\ref{vii4d}) into (\ref{vibb}), we obtain
\bea
\label{S1}
g_{(1)ij}&=& b R_{ij} + c g_{ij} R + \cdots \\
b&=& -\frac{3}{6+\Phi} \nn
c&=& -\frac{3\ \left\{{\Phi'^2}-6\ (\Phi''+8\ V)\right\}}{2\ (6+\Phi)
\ \left\{-2\ {\Phi'^2}+(18+\Phi)\ (\Phi'' +8\ V)\right\}}\ .
\eea
Here $\cdots$ expresses the terms containing the derivative 
with respect to $x_i$. 
Further, substituting (\ref{vii4d}) and (\ref{S1}) into 
(\ref{phi2}), we obtain $\phi_{(2)}$ in terms of $g_{(0)ij}$ 
and $\phi_{(0)}$. 
Substituting (\ref{vii4d}), (\ref{S1}) and the obtained 
expression of $\phi_{(2)}$ into 
(\ref{viii}), one gets the expression of $g^{ij}g_{(2)ij}$. 
Finally  substituting the obtained expressions into the 
expression for the anomaly (\ref{ano}), we obtain,  
\bea
\label{AN1}
S_{\rm ln}&=&{1 \over 16\pi G}\int d^4x \sqrt{-g_{(0)}}\left[
h R_{(0)}^2 + k R_{(0)ij}R_{(0)}^{ij} + \ \cdots \right] \\
h&=& \left[ 3\ \left\{(24-10\ \Phi)\ {\Phi'^6} \right. \right. \nn
&& + \big(62208+22464\ \Phi+2196\ {\Phi^2}+72
\ {\Phi^3}+{\Phi^4}\big)\ \Phi''\ {{(\Phi''+8\ V)}^2} \nn
&& + 2\ {\Phi'^4}\ \left\{\big(108+162\ \Phi+7\ {\Phi^2}\big)\ 
\Phi''+72\ \big(-8+14\ \Phi+{\Phi^2}\big)\ V\right\} \nn
&& - 2\ {\Phi'^2}\ \left\{\big(6912+2736\ \Phi+192
\ {\Phi^2}+{\Phi^3}\big)\ {\Phi''^2} \right. \nn
&& + 4\ \big(11232+6156\ \Phi+552\ {\Phi^2}
+13\ {\Phi^3}\big)\ \Phi''\ V \nn
&& \left. + 32\ \big(-2592+468\ \Phi+96\ {\Phi^2}+5
\ {\Phi^3}\big)\ {V^2}\right\} \nn
&& \left.\left. - 3\ (-24+\Phi)\ {{(6+\Phi)}^2}\ {\Phi'^3}\ (
\Phi'''+8\ V')\right\}\right] \big/  \nn
&& \left[16\ {{(6+\Phi)}^2}\ \left\{-2\ {\Phi'^2}
+(24+\Phi)\ \Phi''\right\}\ \left\{-2\ {\Phi'^2} \right.\right. \nn
&& \left.\left.+(18+\Phi)\ (\Phi''+8\ V)\right\}^2\right]\nn
k &=&-\frac{3\ \left\{(12-5\ \Phi)\ {\Phi'^2}+(288+72\ 
\Phi+{\Phi^2})\ \Phi''\right\}}{8\ {{(6+\Phi)}^2}\ 
\left\{-2\ {\Phi'^2}+(24+\Phi)\ \Phi''\right\}}\ .
\eea
Here $\cdots$ expresses the terms containing the derivative 
with respect to $x_i$. In case of the dilaton gravity in 
\cite{NOano} corresponding to $\Phi=0$ (or more generally 
in case that the axion is 
included as in \cite{NOOSY}), we have the following expression:
\bea
\label{Dxix}
S_{\rm ln}&=&{l^3 \over 16\pi G}\int d^4x \sqrt{-g_{(0)}} 
\left[ {1 \over 8}R_{(0)ij}R_{(0)}^{ij}
-{1 \over 24}R_{(0)}^2 \right. \nn
&& + {1 \over 2} R_{(0)}^{ij}\partial_i\varphi_{(0)}
\partial_j\varphi_{(0)} - {1 \over 6} R_{(0)}g_{(0)}^{ij}
\partial_i\varphi_{(0)}\partial_j\varphi_{(0)}  \nn
&& \left. + {1 \over 4}
\left\{{1 \over \sqrt{-g_{(0)}}} \partial_i\left(\sqrt{-g_{(0)}}
g_{(0)}^{ij}\partial_j\varphi_{(0)} \right)\right\}^2 + {1 \over 3}
\left(g_{(0)}^{ij}\partial_i\varphi_{(0)}\partial_j\varphi_{(0)} 
\right)^2 \right]\ .
\eea
Here $\varphi$ can be regarded as dilaton. When $\Phi$ is 
not trivial, of course, there appear  extra terms which are  
denoted by $\cdots$ in (\ref{AN1}). When $\Phi$ is 
not trivial, for example, the coefficient of 
$\left(g_{(0)}^{ij}\partial_i\varphi_{(0)}\partial_j\varphi_{(0)} 
\right)^2 $ becomes dilaton dependent. And there would appear the 
terms like $R_{(0)}g_{(0)}^{ij}\partial_i\varphi_{(0)}
\partial_j\varphi_{(0)}$ and $R_{(0)}^{ij}\partial_i\varphi_{(0)}
\partial_j\varphi_{(0)}$ and their dilaton dependent coefficients are quite
complicated.

We should also note that the expression (\ref{AN1}) cannot be 
rewritten as a sum of the Gauss-Bonnet 
invariant $\tilde G$ and the square of the Weyl tensor $F$, 
which are 
\bea
\label{GF} 
\tilde G&=&R^2 -4 R_{ij}R^{ij} 
+ R_{ijkl}R^{ijkl} \nn
F&=&{1 \over 3}R^2 -2 R_{ij}R^{ij} 
+ R_{ijkl}R^{ijkl} \ ,
\eea
This is the signal that the conformal symmetry is broken. 
In the limit of $\Phi\rightarrow 0$, we obtain 
\bea
\label{Lmt}
h&\rightarrow& {3\cdot 62208 \Phi'' (8V)^2 \over 16\cdot 6^2
\cdot 24 \cdot 18^2 \Phi'' (8V)^2} = {1 \over 24} \nn
k&\rightarrow& - {3\cdot 288 \Phi'' \over 8\cdot 6^2\cdot 24 
\Phi''}=-{1 \over 8}
\eea
and we can find that the standard result (conformal 
anomaly of ${\cal N}=4$ super YM theory) is reproduced. 
In order that the region near the boundary at $\rho=0$ is 
asymptotically AdS, we need to require $\Phi\rightarrow 0$ 
and $\Phi'\rightarrow 0$ when $\rho \rightarrow 0$. 
We can also confirm that $h\rightarrow {1 \over 24}$ and 
$k\rightarrow -{1 \over 8}$ in the limit of $\Phi\rightarrow 0$ 
and $\Phi'\rightarrow 0$ 
even if $\Phi''\neq 0$ and $\Phi'''\neq 0$. 
In the AdS/CFT correspondence, $k$ and $h$ should be related with 
the central charge $c$ of the conformal field theory 
(or its analog for non-conformal theory). Since 
we have two functions $h$ and $k$, there are two natural ways to define 
the c-function when the conformal field theory is deformed: 
\be
\label{CC}
c_1={24\pi h \over G}\ ,\quad 
c_2=-{8\pi k \over G}\ .
\ee
Note that above can be considered as auxiliar consequence of the found 
conformal anomaly.
If we put $V(\phi)=4\lambda^2 + \Phi(\phi)$, we have 
$l=\left(12\over V(0)\right)^{1 \over 2}$. We should note that 
we have chosen $l=1$ in the expressions in (\ref{CC}). We can 
restore $l$-dependence by changing $h\rightarrow l^3 h$ and $k\rightarrow 
l^3 k$ and $\Phi'\rightarrow l\Phi'$, $\Phi''\rightarrow 
l^2\Phi''$ and $\Phi''' \rightarrow l^3\Phi'''$ in (\ref{AN1}). 
Then in the limit of $\Phi\rightarrow 0$, we obtain
\be
\label{CCl}
c_1\ ,\quad c_2\ \rightarrow {\pi \over G}
\left(12\over V(0)\right)^{3 \over 2}\ ,
\ee
which agrees with the definition used in the  
works \cite{GPPZ2, FGPW2} (where equations of motion were actually used) in
above limit. 
The $c_1$- or $c_2$-functions in (\ref{CC}) give the new candidate
for c-function away of conformity. 

The definitions of the c-functions in (\ref{d2c}) and (\ref{CC}), 
are, however, not always good ones since our results are too wide. 
That is, we have obtained the conformal anomaly for arbitrary 
dilatonic background which may not be the solution of original 
$d=5$ gauged supergravity. As only solutions of d5 gauged supergravity 
describe RG flows of dual QFT it is not strange that above candidate 
c-functions are not acceptable. They quickly become non-monotonic 
and even singular in explicit examples. In such situation
(taking into account the construction method) they presumbly 
measure the
deviations from SG description and should not be taken seriously. 
As pointed in \cite{MTR}, it might be necessary to impose the 
condition $\Phi'=0$ on the boundary which follows 
from the equations of motion of d5 gauged SG. 
Anyway as $\Phi'= 0$ on the boundary in the solution which has 
the asymptotic AdS region, we can add any function 
which proportional to the power of $\Phi'= 0$ to the previous 
expressions of the c-functions in (\ref{d2c}) and (\ref{CC}). 
As a trial, if we put $\Phi'=0$, we obtain 
\bea
\label{d2cb}
c&=&{3 \over 2G}\left[ {l \over 2} + {1 \over l} 
{1 \over \Phi(\phi_{(0)})  +{2 \over l^2 }}\right]
\eea
instead of (\ref{d2c}) and
\bea
\label{CCb}
c_1&=&{9\pi \over 2G}{62208+22464\Phi
+2196 \Phi^2 +72 \Phi^3+ \Phi^4 \over 
(6+\Phi)^2(24+\Phi)(18+\Phi)^2} \nn
c_2&=&{3\pi \over G}{288+72 \Phi+ \Phi^2 \over 
(6+\Phi)^2(24+\Phi)}
\eea
instead of (\ref{CC}). \footnote{Note that our proposal is clearly more 
restrictive than the c-function of ref.\cite{GPPZ} because it works only 
in situation when AdS/CFT correspondence is valid. The correspondent 
trace anomaly was obtained under such assumption. In particular,
above c-functions should not be considered for negative potentials.} We 
should note that there disappear the 
higher derivative terms  like $\Phi''$ or $\Phi'''$. That will be our 
final proposal for acceptable c-function in terms of dilatonic potential. 
The given c-functions in (\ref{CCb}) also have the property 
(\ref{CCl}) and reproduce the known result for the central charge 
on the boundary. 
Since $\Phi'\rightarrow 0$ in the asymptotically AdS region 
even if the region is UV or IR, the given c-functions in 
(\ref{d2cb}) and (\ref{CCb}) have fixed points in the 
asymptotic AdS region ${d c \over dU}={dc \over d\Phi}
{d\Phi \over d\phi}{d\phi \over dU}\rightarrow 0$, where 
$U=\rho^{-{1 \over 2}}$ is the radius coordinate in AdS 
or the energy scale of the boundary field theory.

We can now check the monotonity in the c-functions. 
For this purpose, we consider some examples. 
In \cite{FGPW} and \cite{GPPZ}, the 
following dilaton potentials appeared:
\bea
\label{FGPWpot}
4\lambda^2 + \Phi_{\rm FGPW}(\phi)
&=&4\left(\exp\left[ \left({4\phi \over \sqrt{6}}\right)
\right] + 2 \exp\left[ -\left({2\phi \over \sqrt{6}}\right)
\right]\right)\\
\label{GPPZpot}
4\lambda^2 + \Phi_{\rm GPPZ}(\phi)
&=&{3 \over 2}\left(3+\left(\cosh\left[ \left({2\phi \over 
\sqrt{3}}\right)\right]\right)^2 + 4\cosh\left[ \left(
{2\phi \over \sqrt{3}}\right)\right]\right) \ . 
\eea
In both cases $V$ is a constant and $V=-2$. 
In the classical solutions for the both cases, 
$\phi$ is the monotonically decreasing 
function of the energy scale $U= \rho^{-{1 \over 2}}$ and 
$\phi=0$ at the UV limit corresponding 
to the boundary. 
Then in order to know the energy scale dependences 
of $c_1$ and $c_2$, we only need to investigate the $\phi$ 
dependences of $c_1$ and $c_2$ in (\ref{CCb}). As the potentials 
and also $\Phi$ have a minimum $\Phi=0$ 
at $\phi=0$, which corresponds to the UV boundary in the solutions 
in \cite{FGPW} and \cite{GPPZ}, and $\Phi$ is monotonicaly 
increasing function of the absolute value $|\phi|$, we only 
need to check the monotonities of $c_1$ and $c_2$ with respect 
to $\Phi$ when $\Phi\geq 0$. From (\ref{CCb}), we find
\bea
\label{monot}
&& {d \left(\ln c_1\right) \over d\Phi} \nn
&& = - {18\left(622080 + 383616\Phi + 64296\Phi^2 + 4548\Phi^3 
+ 130\Phi^4 + \Phi^5\right) \over 
(6 + \Phi)(18 + \Phi) (24 + \Phi) (62208+22464\Phi
+2196 \Phi^2 +72 \Phi^3+ \Phi^4 )} \nn
&& <0 \nn
&& {d \left(\ln c_2\right) \over d\Phi}=
 - {5184 + 2304\Phi + 138\Phi^2 + \Phi^3 
 \over (6 + \Phi) (24 + \Phi) (288+72 \Phi+ \Phi^2)}
<0 \ .
\eea
Therefore the c-functions $c_1$ and $c_2$ are monotonically 
decreasing function of $\Phi$ or increasing function of the 
energy scale $U$ as the c-function in \cite{DF,GPPZ}. 
We should also note that the 
c-functions $c_1$ and $c_2$ are positive definite for 
non-negative $\Phi$.  For $c$ in (\ref{d2cb}) for $d=2$ 
case, it is very straightforward to check the monotonity and the 
positivity. 

In \cite{GPPZ2}, another c-function has been proposed 
in terms of the metric as follows:
\be
\label{gppzC}
c_{\rm GPPZ}=\left({d\varphi \over d y}\right)^{-3}\ , 
\ee
where the metric is given by 
\be
\label{gppzC2}
ds^2=dy^2 + \e^{2\varphi}dx_\mu dx^\mu\ .
\ee
The c-function (\ref{gppzC}) is positive and 
has a fixed point in the 
asymptotically AdS region again and the c-function is also 
 monotonically decreasing function of the energy scale. 
The c-functions (\ref{d2cb}) and (\ref{CCb}) proposed 
in this paper are given in terms of the dilaton potential, 
not in terms of metric, but it might be interesting that 
the c-functions in (\ref{d2cb}) and (\ref{CCb}) have the 
similar properties (positivity, monotonity and fixed point 
in the asymptotically AdS region).

In summary, we found the conformal anomaly from d3 and d5 gauged 
supergravity with single scalar and arbitrary scalar potential 
on the scalar-gravitational background. It corresponds to 
the conformal anomaly of dual boundary
theory. The proposal to define c-function which is positively definite and
monotonically decreasing and defined in terms of dilatonic potential is given.
Our work may be extended for d5 gauged SG with bigger number 
of scalars (say ${\cal N}=8$ gauged SG) and arbitrary 
scalar potential. The final result appears 
in really complicated and lengthy form as it will be shown 
in another place.
This opens the possibility of explicit check if the results on 
RG flows 
in dual gauge theory (deformed ${\cal N}=4$ super Yang-Mills) 
presented in refs.\cite{DF,GPPZ} from bulk side indeed 
describe 4d gauge Yang-Mills 
theory with lesser supersymmetry and the correspondent 
identification is
correct. From another side, our conformal anomaly in the spirit of 
ref.\cite{myers} may be used to calculate the Casimir energy 
in dilatonic gravity.
As the final remark let us note that dilaton-dependent conformal 
anomaly found in this work may be used for calculation of anomaly 
induced effective action of non-conformal boundary QFT 
in the presence of scalars (see ref.\cite{brevik} 
for related example of dilaton dependent 
induced effective action in SUSY Yang-Mills theory).

\end{document}